\documentclass[twoside,fleqn]{ActaStyle}
\usepackage{times,cite}

\usepackage{lscape}             
\usepackage{graphicx,epsfig}    
\usepackage[tbtags]{amsmath}    

\usepackage{amssymb}
\usepackage{pslatex}

\newcommand{\sqrtsnn}{\sqrt{s_{_{NN}}}}

\newcommand{\jpsi}{J/\psi}
\newcommand{\pom}{I\!\!P}
\newcommand{\gaga}{\gamma\,\gamma}
\newcommand{\gp}{\gamma\,p}
\newcommand{\gA}{\gamma\,A}

\def\authorheading{David d'Enterria for the PHENIX Collaboration}

\def\shorttitle{Photoproduction of $J/\psi$ and high-mass $e^+e^-$
in UPC Au+Au at $\sqrtsnn$ = 200 GeV}

\begin{document}

\pagerange{1}{6}

\title{%
COHERENT PHOTOPRODUCTION OF $J/\Psi$ AND HIGH-MASS $e^+e^-$ PAIRS\\
IN ULTRA-PERIPHERAL AU+AU COLLISIONS AT $\sqrtsnn$ = 200 GeV
}

\author{David d'Enterria\\ 
for the PHENIX Collaboration}
{Nevis Labs, Columbia University, Irvington, NY 10533, and New York, NY 10027, USA\\
LLR, Ecole Polytechnique, 91128 Palaiseau, France, EU}

\day{Dec 30, 2005}

\abstract{%
High energy ultra-peripheral collisions (UPC) of heavy-ions generate
strong electromagnetic fields which open the possibility to study 
$\gaga$ and $\gamma$-nucleus processes in a kinematic regime so far 
unexplored. We report on preliminary PHENIX results of $J/\psi$ 
and high-mass $e^+e^-$ photoproduction at mid-rapidity in coherent 
electromagnetic Au+Au interactions at $\sqrtsnn$ = 200 GeV tagged 
with forward neutron emission from the $Au^\star$ dissociation.
}
\pacs{%
13.40.-f, 13.60.-r, 24.85.+p, 25.20.-x, 25.20.Lj, 25.75.-q
}

%
\section{INTRODUCTION}
\label{sec:intr} \setcounter{section}{1}\setcounter{equation}{0}

\subsection{$\gaga$, $\gp$ physics at $e^+e^-$ and $e p$ colliders}

\begin{figure}[htbp]
\begin{center}
\includegraphics[height=6.5cm]{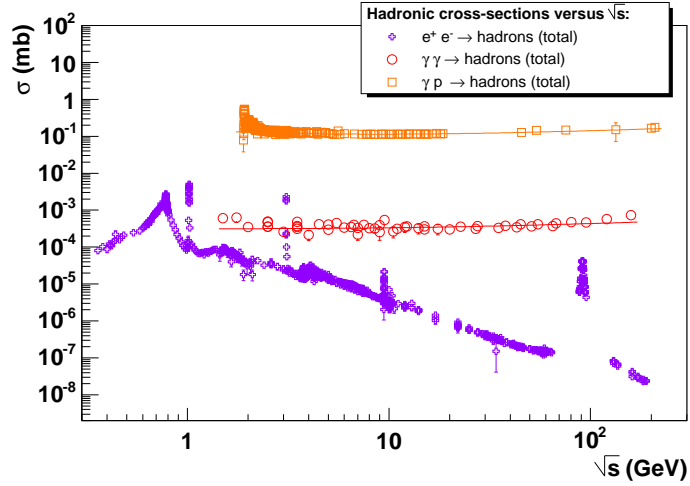}
\end{center}
\vspace{-0.5cm}
\caption{Total hadronic cross-section in $e^+e^-$, $\gaga$ and $\gp$ collisions
as a function of the center-of-mass energy $\sqrt{s}$. The data is from the PDG~\protect\cite{pdg}, 
and the fits to $\gaga$, $\gp$ data are the latest Regge-theory parametrizations 
of the COMPETE collaboration~\cite{compete}.}
\label{fig:ee_gg_gq_sigma_had}
\end{figure}

\begin{figure}[htbp]
\begin{center}
\includegraphics[height=3.8cm]{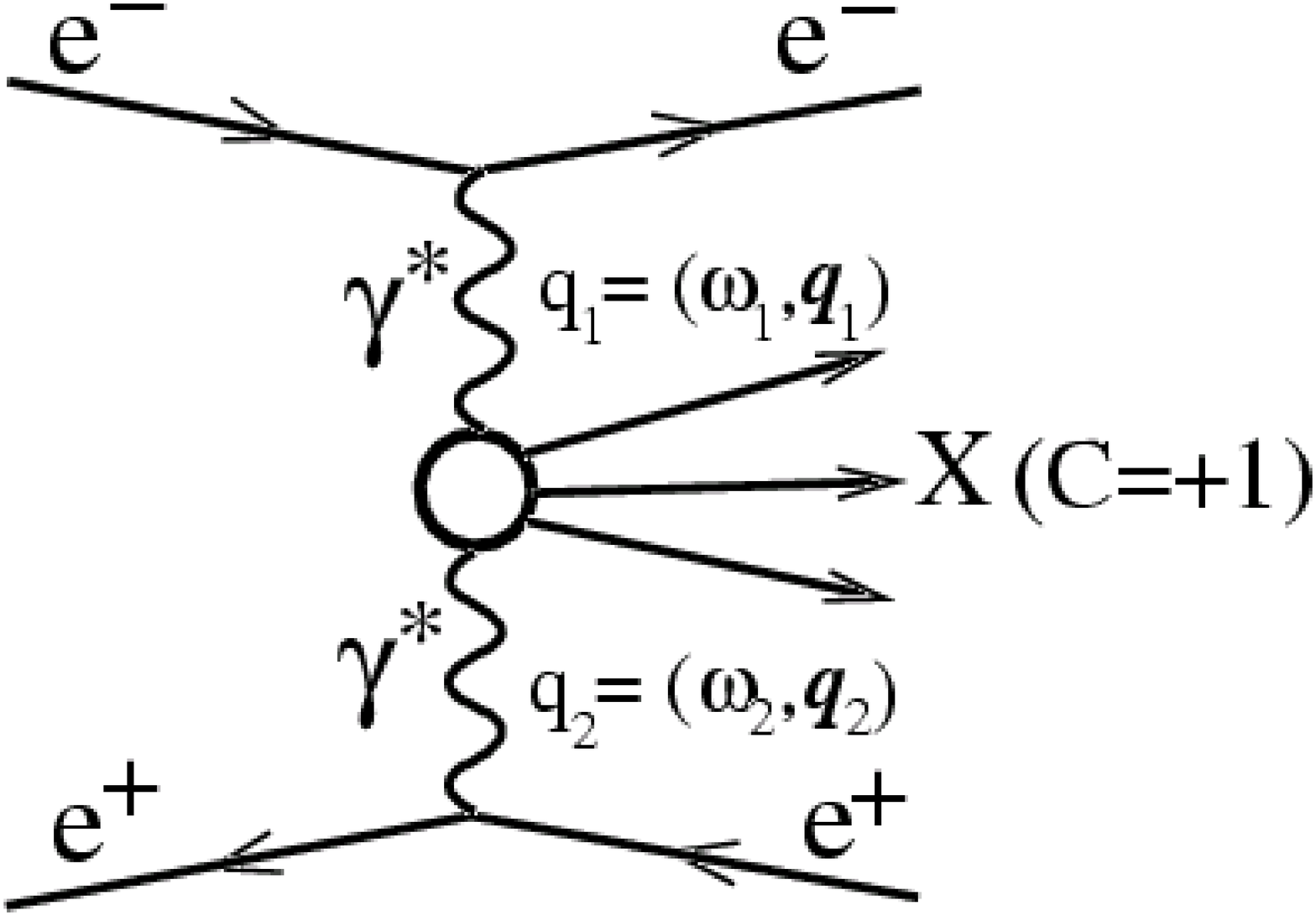}
\includegraphics[height=4.cm]{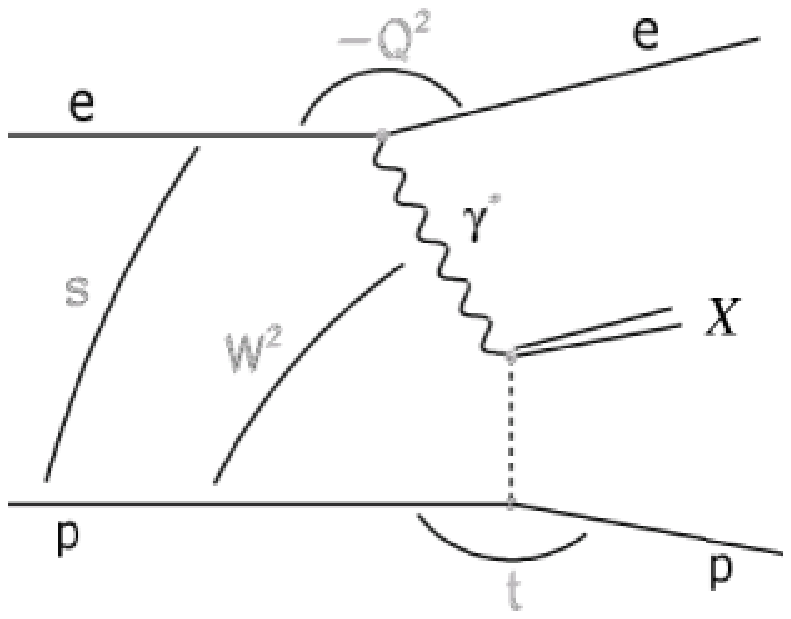}
\end{center}
\vspace{-0.5cm}
\caption{Typical diagrams for photon-photon (left) and photon-proton (right) 
reactions at $e^+e^-$ and $e p$ colliders respectively.}
\label{fig:diag_gg_gp}
\end{figure}

Our knowledge about elementary particles and their fundamental interactions is 
mainly obtained through the study of high energy particle collisions. 
A part from the more ``conventional'' $e^+e^-$, $e p$ (DIS) and $pp$, $\bar{p}p$ collisions, 
high-energy $\gaga$~\cite{budnev74} and $\gp$~\cite{schuler93} processes provide 
an interesting and complementary approach to study strong and electro-weak interactions. 
At high energies, a (quasi) real photon can interact as a point-like particle (e.g. Compton-like 
scattering) or through quantum fluctuations into 
a charged-fermion pair ($q\bar{q}$, $l^+l^-$), a charged-boson pair (e.g. $W^+W^-$) or 
directly into a 
vector meson (carrying the same quantum numbers $J^{PC} = 1^{-\,-}$ as the photon). 
At energies below the $W^+W^-$ threshold, the photon wave function can be 
written as~\cite{schuler93}
\begin{equation}
|\gamma\rangle = c_0\,|\gamma_0\rangle \;\;+ \sum_{V=\rho,\omega,\phi,J/\psi,\Upsilon} c_V\,|V\rangle 
\;+ \sum_{q=u,d,s,c,b} c_q\,|q\bar{q}\rangle \;+ \sum_{l=e,\mu,\tau} c_l\,|l^+l^-\rangle .
\label{eq:photon}
\end{equation}
Though in practice the first term dominates (i.e. $c_0\approx$ 1), hadronic/partonic fluctuations 
(under the non-perturbative ``Vector-Dominance-Model'' form $\gamma\rightarrow V$ or, at 
larger $q\bar{q}$ virtualities, via ``resolved'' quark pairs $\gamma\rightarrow q\bar{q}$
which can, at their turn, radiate also gluons) interact strongly and give the largest 
contribution to the total $\gaga$, $\gp$ cross-sections at high energies. 
Fig.~\ref{fig:ee_gg_gq_sigma_had} shows the total hadronic cross-sections measured in
$e^+e^-$, $\gaga$ and $\gp$ collisions as a function of the center-of-mass energy 
$\sqrt{s}$. While, the ($s$-channel) $e^+e^-$ annihilation cross-section 
decreases with increasing energy ($\sigma_{ee\rightarrow hadrons}\propto \alpha_{em}^2/s$), 
photon-induced collisions behave like $pp$, $p\bar{p}$ collisions. Their cross-sections
rise monotonically with $\sqrt{s}$ in agreement with an increasingly large Pomeron 
(or two-gluon) exchange contribution as described by Regge 
phenomenology~\cite{donnachie_landshoff92,compete}.
\noindent
Although in the present-day there are no beams of high energy {\it real} photons 
available, beams of high energy {\it quasi-real} or {\it virtual} photons are commonly
generated at $e^+e^-$ and $e p$ colliders. Indeed, the electromagnetic field of any 
relativistic charged particle can be described as a flux of ``equivalent'' photons with 
energy distribution $f_{\gamma/e}(z) = dN_\gamma/d\omega$ given by the Weizs\"acker-Williams~\cite{weizsacker_williams} 
or Equivalent Photon Approximation (EPA) formula, which for an $e^\pm$ beam reads:
\begin{equation}
\frac{dN_\gamma}{dz} \approx \frac{\alpha_{em}}{2\pi}\frac{1}{z}[1+(1-z)^2]\ln\frac{Q_{max}^2}{Q_{min}^2}\;\;\;\;,
\mbox{ with }\;\;\;z=\omega/E_e\,.
\label{eq:ww}
\end{equation}
Here $z$ is the fraction of the $e^\pm$ energy carried by the photon and
$Q^2=-(q^2)$ the $\gamma$ virtuality. $Q_{min}^2 = m_e^2 z^2/(1-z)\approx 0$ GeV$^2/c^2$, 
whereas $Q^2_{max}$ depends on the properties of the produced $X$, e.g. 
$Q^2_{max} \approx m_\rho^2$ for hadron production, and $Q^2_{max} \approx W^2$ 
for $X = l^+l^-$, where $W^2=W_{\gaga}^2=4\omega_1\omega_2$ 
(or $W^2=W^2_{\gp} = 4E_p\,E_e\,z$) is the squared $\gaga$ (or $\gp$) 
c.m. energy. Fig.~\ref{fig:diag_gg_gp} shows the typical two-photon 
($e^+e^-\rightarrow e^+e^-\gaga\rightarrow e^+e^- X$) and photon-proton 
($e p\rightarrow e\,\gp\rightarrow e p X $) processes commonly studied at LEP 
and HERA colliders. Two basic properties characterize $\gaga$ and $\gp$
collisions:
\begin{itemize}
\item The variable-energy photon ``beam'' always carries less energy than the parent 
lepton beam (the spectrum (\ref{eq:ww}) has a soft bremsstrahlung shape, 
$dN_\gamma/d\omega\propto ln(1/\omega)/\omega$, peaked at low $\omega$) 
and thus the available center-of-mass energies are typically $0.1\sqrt{s} \lesssim W_{\gaga,\gp}\lesssim0.5\sqrt{s}$.
\item The (almost) real photons beams have, by definition, $Q^2\approx$ 0 (they are 
often kinematically selected by requiring the scattered parent $e^\pm$ to be close to 
the incident direction) and thus the produced $X$ have always small transverse momentum
(at variance with DIS scattering where the $\gamma^\star$ has 
$Q^2\gtrsim$ 2 GeV$^2/c^2$ and the produced particles have large $p_T$).
\end{itemize} 

\noindent
The main interest of $\gamma$-induced processes at $e^+e^-,ep$ colliders so far is 
that they allow for precision studies of QCD dynamics (tests of Regge theory, low-energy quark model
spectroscopy, BFKL evolution, hard-scattering factorization, diffractive interactions, ...) 
in a background-free environment and with a comparatively simpler initial state than in 
hadron-hadron collisions. QED processes (e.g. $l^+l^-$ production), on the other hand, 
have been less addressed than QCD-related studies. Typical physics measurements 
encompass~\cite{photoprod_rep05}: 
(i) total hadronic cross-sections, (ii) $\gaga$ widths ($\Gamma_{\gaga}$) of $C$-even 
(scalar $0^{-\,+,++}$ and tensor $2^{++,...}$ ) mesons, 
(iii) double vector meson production ($\rho,\omega,\phi,\jpsi,\Upsilon$),
(iv) structure function of the photon, 
(v) (hard) photo-production of quarkonia ($\jpsi,\Upsilon$), open heavy quarks 
and jets; and (vi) diffractive structure functions.
The corresponding two-photon ($\sigma_{\gaga\rightarrow X}$) or photo-production 
($\sigma_{\gp\rightarrow X}$) cross-sections are theoretically computed using 
different methods depending on the c.m. energy $W$ and the nature of the produced particle 
$X$. Regge-based models (as e.g. in the DPM~\cite{dpm}), (Generalized) Vector-Meson-Dominance 
(VDM) approaches~\cite{vdm}, pQCD factorization~\cite{klasen02} (with different available 
parametrizations of the parton distribution function of the photon~\cite{aurenche05}), 
color-dipole (or saturation) approaches~\cite{teaney03,goncalves},
or a combination (of some) of them (as e.g. in PHOJET~\cite{phojet} or 
PYTHIA~\cite{schuler93} event generators) are often used.\\

\noindent
At future linear collider facilities using beams of back-scattered 
laser photons off beam electrons, new prospects appear both within the
Standard Model (Higgs production, triple gauge couplings, precision top physics) 
and beyond-SM (SUSY particle production, extra-dimensions, ...)~\cite{deroeck04}.

%
\subsection{$\gaga$, $\gA$ physics in ultra-peripheral A+A reactions}

As aforementioned, any relativistic charged particle generates a flux of
quasi-real photons which can be used for photoproduction studies.
Though in the previous Section we have considered the case of $e^\pm$ beams, 
the same holds obviously true for accelerated protons~\cite{klein_nystrand04} 
or nuclei~\cite{baur98,baur01,bertula05}. 
The corresponding Equivalent Photon Approximation formula (\ref{eq:ww})
for an extended object such as a proton or a nucleus with charge $Z$ and mass $m_{A}$ 
is~\cite{budnev74}
\begin{equation}
f_{\gamma/A}(z) = \frac{\alpha_{em}\,Z^2}{2\pi}\frac{1+(1-z)^2}{z}
\int_{Q^2_{min}}^\infty\frac{Q^2-Q_{min}^2}{Q^4}\,|F(Q^{2})|^2dQ^2,
\label{eq:ww_Z}
\end{equation}
where $Q^2$ is the momentum transfer from the projectile, $Q_{min}^2 = m_A^2 z^2/(1-z)$,
and $F(Q^2)$ a nuclear form factor describing its spatial distribution.
From Eq.~(\ref{eq:ww_Z}) one can see that 
the equivalent photon flux in relativistic heavy-ion collisions (with charge Z$\sim$ 80 
for conventional Au or Pb beams) has an amplification factor of $Z^2\sim$ 6.5$\cdot 10^3$ 
compared to $e^\pm$ or proton beams. The corresponding $\gaga$ cross-sections are
a factor $Z^4\sim$ 4$\cdot 10^7$ larger ! This fact drives the main
interest on coherent electromagnetic (or ultra-peripheral) interactions of 
heavy-ions~\cite{baur98,baur01,bertula05}.
For collisions between heavy-ions, rather than Eq.~(\ref{eq:ww_Z}) it is more useful 
the expression for the equivalent photon spectrum above a given minimum impact 
parameter~\cite{jackson} which is usually chosen as twice the nuclear radius, 
$b > b_{min} \approx 2\,R_A$ (i.e. $b_{min}\approx$ 15 fm for Au, Pb beams), 
so that the contribution from non-electromagnetic ion-ion interactions can be removed:
\begin{equation}
f_{\gamma/A}(z,b>b_{min}) = \frac{\alpha_{em}\,Z^2}{\pi}\frac{1}{z}
\left[2x\,K_0(x)K_1(x)-x^2\left(K_1^2(x)-K_0^2(x)\right)\right],
\label{eq:ww_bZ}
\end{equation}
where $K_{0,1}$ are the modified Bessel functions of the 2nd kind and $x=z\,m_A\,b_{min}$.
The validity of the application of the EPA formula for heavy-ions 
is limited to the case where {\it all} the protons of the nucleus interact 
electromagnetically in a coherent way. In that case, the wavelength of the resulting 
photon is larger than the size of the nucleus, given by its radius $R_A$.
This ``coherence'' condition limits the maximum virtuality of the produced photon 
to very low values~\cite{baur98}:
\begin{equation}
Q^2=\left(\omega^2/\gamma^2+q_\perp^2\right)\lesssim 1/R_A^2\;\;\;\mbox{ (where $\gamma$ is the beam Lorentz factor)},
\label{eq:Q2_coh}
\end{equation}
and thus for most purposes (but for lepton pair production) those photons 
can be considered as (quasi) real with maximum energy and perpendicular 
momentum:
\begin{equation}
\omega<\omega_{max} \approx \frac{\gamma}{R}\;\;\;\mbox{ , and } 
q_\perp\lesssim\frac{1}{R}\approx 30 \;\mbox{MeV}.
\label{eq:omega_max}
\end{equation}
At RHIC (LHC) energies with $\gamma$ = 100 ($\gamma\approx$ 2800 for Pb), 
the maximum photon energy in the lab system is $\omega_{max}\approx$ 3 GeV (80 GeV).
Thus, the corresponding maximum energies in the c.m. system for 
$\gaga$ and $\gA$ processes are $W_{\gaga}^{max}\approx$ 6 (160) GeV 
and $W^{max}_{\gA}\approx$ 34 (940) GeV respectively.

\subsection{Production cross-sections in UPC collisions}

\begin{figure}[htbp]
\begin{center}
\includegraphics[height=4.cm]{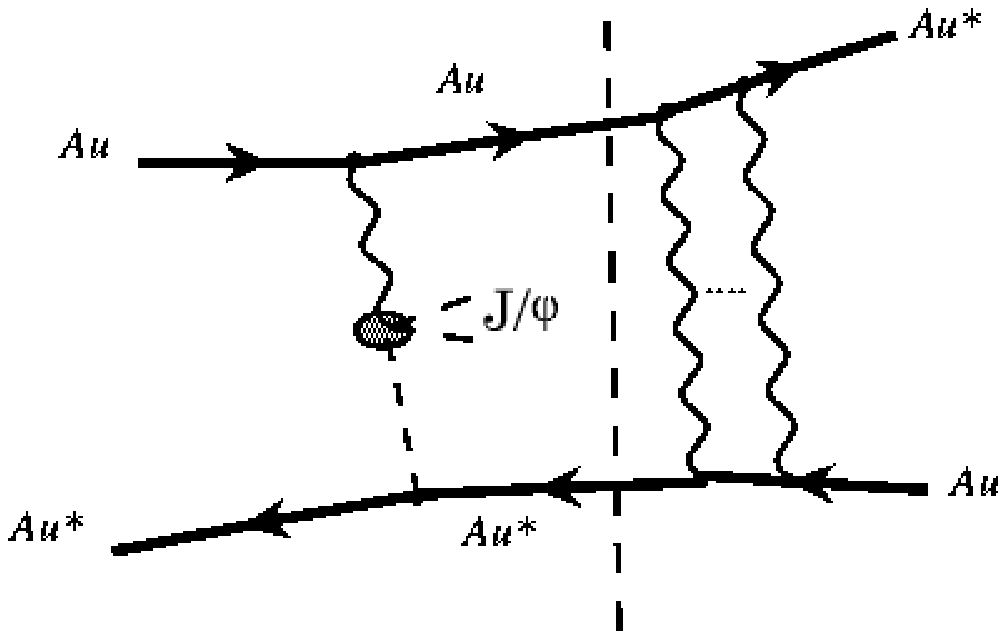}
\includegraphics[height=4.5cm]{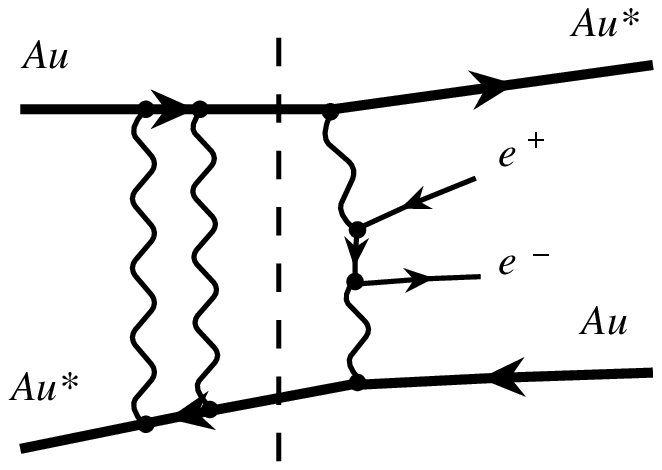}
\end{center}
\vspace{-0.5cm}
\caption{Lowest order Feynman diagrams for $\jpsi$ (left) and dielectron pair (right) 
production in $\gaga$ and $\gA$ processes accompanied by Au Coulomb 
excitation in ultra-peripheral Au+Au collisions.}
\label{fig:diag_gg_gA}
\end{figure}

\noindent
The production cross-section of a system $X$ in UPC A+A collisions is computed as 
the convolution of the corresponding equivalent photon spectrum (\ref{eq:ww_bZ}) 
with the elementary photonuclear or two-photon cross-sections:
\begin{eqnarray}
\sigma(A+A\rightarrow \gamma +A+A\rightarrow A+A+X) & = &
\int_{0}^{1}f_{\gamma/A}(z)\sigma_{\gA\rightarrow X}(W_{\gA})dz\\
\sigma(A+A\rightarrow \gaga +A+A\rightarrow A+A+X) & = &
\int_{0}^{1}\int_{0}^{1}f_{\gamma/A}(z_1)f_{\gamma/A}(z_2)\sigma_{\gaga\rightarrow X}(W_{\gaga})dz_1 dz_2\;,
\label{eq:sigma_AA}
\end{eqnarray}
where, as aforementioned, the elementary $\sigma_{\gA\rightarrow X}$ and
$\sigma_{\gaga\rightarrow X}$ are theoretically computed using different methods 
depending on the c.m. energy $W$ and the particle $X$. The two cases of interest 
in this paper, shown in Figure~\ref{fig:diag_gg_gA}, are the coherent photoproduction of:
\begin{description}
\item I. $\jpsi$, the heaviest vector meson effectively accessible in $\gA$ 
collisions at RHIC, via:\\ 
\noindent $A+A \;(\rightarrow \gamma+A) \rightarrow A^\star+A^{(\star)}+\jpsi$.
\item II. High mass di-electron continuum in $\gaga$ collisions:
$A+A \; (\rightarrow \gamma+\gamma) \rightarrow A^\star+A^{(\star)}+e^++e^-$.
\end{description}
For the first process (I), $\sigma_{\gA\rightarrow \jpsi A}$ has been determined 
using perturbative QCD for the $\gp\rightarrow\jpsi\,p$ process~\cite{strikman95}
plus initial-state shadowing and final-state $\jpsi$-nucleus interaction~\cite{strikman05}, 
or within the color-dipole formalism~\cite{goncalves}. The cross-section for heavy vector 
meson ($\jpsi,\Upsilon$) photoproduction is found to depend 
(i) {\it quadratically} on the gluon density $G_A(x,Q^2)$~\cite{ryskin95}:
\begin{equation}
\left .\frac{d\sigma(\gA\rightarrow V\,A)}{dt}\right|_{t=0} = 
\frac{\alpha_s^2\Gamma_{ee}}{3\alpha M_V^5}16\pi^3\left[xG_A(x,Q^2)\right]^2
\;\;\mbox{, with }\;Q^2=M_V^2/4\;\;\mbox{, and }\;x=M_V^2/W_{\gA}^2,
\end{equation}
as well as on (ii) the probability of rescattering or absorption of the $Q\overline{Q}$ 
pair as it traverses the nucleus. The study of quarkonia production in 
$\gA$ collisions at RHIC or LHC energies is, thus, considered an excellent 
probe of (i) the gluon distribution function $G_A(x,Q^2)$ in nuclei, and
(ii) vector-meson dynamics in nuclear matter.\\

\noindent
For the second purely electromagnetic process (II), the total
$\sigma_{\gaga \rightarrow \gaga e^+e^-}$ cross-section is huge 
(about 30 kb for Au+Au at RHIC !) but strongly peaked at forward-backward 
rapidities and at low $m_{inv}\lesssim$ 10 MeV/$c^2$~\cite{baur98}. 
Several calculations exist for multiple $e^\pm$ pair production 
within QED\cite{baur98} 
or solving the semi-classical Dirac equation~\cite{baltz_ee}. 
The main interest of such coherent $\gaga$ collisions is that one 
can test QED in a very strong field regime ($Z\alpha_{em}\approx$ 0.6) 
where perturbative calculations are expected to break~\cite{bertula_baur}.

%
\section{EXPERIMENTAL}

At RHIC energies, the production cross section for vector mesons ($\rho$, $\phi$, 
$\jpsi$, $\Upsilon$) in UPC Au+Au is large and accounts for 
as much as $\sim$10\% of the total $\sigma_{AuAu}\approx$ 7 b nuclear cross 
section~\cite{starlight1,starlight2}. Measurements of coherent photonuclear 
production of the $\rho$ meson~\cite{star_rho} as well as $\gaga$ production 
of {\it low-mass} $e^\pm$ pairs~\cite{star_ee} have been performed by the 
STAR collaboration. The PHENIX analysis presented here 
aims at similar measurements but at higher invariant masses.\\

\noindent
Detailed knowledge of the experimental signatures of coherent UPC events is a basic
pre-requisite to setup an efficient UPC trigger and to define the reconstruction and 
analysis cuts used in the present work. The typical characteristics of UPC Au+Au
events are:
\begin{enumerate}
\item Low central multiplicities: typical values are (well) below $\sim$15 particles. 
\item Low total transverse momentum (``coherence condition''): $p_T < 2\hbar c/R \approx$ 50 MeV/$c$ or 
$p_T\sim m_{inv}/\gamma\approx$ 30 MeV/$c$.
\item Large probability of multiple electromagnetic interactions ($\gamma$-exchanges
shown in Fig.~\ref{fig:diag_gg_gA}) leading to (single or double) nucleus Giant-Dipole-Resonance 
(GDR) excitation followed by $Xn$ neutron(s) decay. 
Typical probabilities are $P_{1n}\sim$30-50\% ($\jpsi$) or $P_{1n}\sim$20\% ($\rho$) 
which factorize out when determining the UPC cross-sections~\cite{bertula_baur,baur03}.
\item Zero net charge: even number of charged tracks of opposite signs.
\item Narrow $dN/dy$ peaked at mid-rapidity (narrower for larger $m_{inv}$).
\end{enumerate}

\noindent
Property 3. is the most useful for UPC tagging and trigger purposes.

%
\subsection{PHENIX setup and luminosity}

The data presented here were collected with the PHENIX detector at BNL RHIC during the 
2004 high luminosity Au+Au run at $\sqrtsnn$ = 200 GeV (Run-4). The gold beams had 
45 or 56 bunches with 10$^9$ ions/bunch and 106 ns crossing-time. Typical 
luminosities at the beginning of the store were $\sim$2 10$^{26}$ cm$^{-2}$s$^{-1}$, 
twice larger than the design luminosity (thanks mainly to an improved vacuum 
system)~\cite{fischer04}.
The PHENIX detector~\cite{phenix_nim} is specifically designed to measure hard probes 
by combining good mass and particle identification (PID) resolution in a broad momentum 
range, high rate capability, and small granularity. 
The central-arm detectors used in this analysis (DC, PC, RICH and EMCal) are those 
needed for the measurement of $\jpsi$ ($e^+e^-$ decay mode) and 
high-mass ($m_{inv}>$ 2 GeV/$c^2$) dielectron pairs at $y$ = 0. 
The ultra-peripheral Au+Au events were tagged by neutron detection 
at small forward angles in the Zero-Degree-Calorimeters (ZDC).\\

\noindent
The momentum and trajectory of the tracks were reconstructed with the central tracking 
(CNT) system (covering $\Delta\eta = 0.7$ and $\Delta \phi = \pi$) based on a 
multi-layer drift chamber (DC) followed by multi-wire proportional chambers 
(PC) with pixel-pad readout, both placed in an axial magnetic field parallel to the beam 
($B_{max}$ = 1.15 T m). 
Electron-positron identification was done with the Ring-Imaging- \v{C}erenkov 
(RICH, with CO$_{2}$ gas radiator) and electromagnetic calorimeter EMCal~\cite{nim_emcal}, 
with a total total solid angle at mid-rapidity of $\Delta\eta \approx 0.7$ and 
$\Delta \phi = \pi$. The PHENIX EMCal consists of six sectors of lead-scintillator 
sandwich calorimeter (PbSc, 15552 individual towers with 5.54 cm $\times$ 5.54 cm 
$\times$ 37.5 cm, 18$X_0$) and two sectors of lead-glass \v{C}erenkov calorimeter 
(PbGl, 9216 modules with 4 cm $\times$ 4 cm $\times$ 40 cm, 14.4$X_0$), at a radial 
distance of $\sim$5\,m from the beam line. 
The ZDC~\cite{nim_zdc,chiu01} hadronic calorimeters 
placed 18~m up- and down- stream of the interaction point, 
cover the very-forward region, $|\theta|<$ 2 mrad, 
and measure the energy of the neutrons coming from the Au$^\star$ Coulomb 
dissociation, with $\sim$20\% resolution.\\

\noindent
The total equivalent sampled luminosity for this study was: 
${\cal L}_{int} = N_{\ensuremath{\it BBC-LL1}}/\sigma_{AuAu} \times \epsilon_{BBC} = \; 120 \pm 10 \;\mu\mbox{b}^{-1}$,
where $N_{\ensuremath{\it BBC-LL1}} = 886\cdot 10^{6}$ is the total number of events 
collected (after vertex cut and QA) by the Beam-Beam-Counter (BBC) minimum bias  Au+Au 
trigger, and we use for this preliminary analysis the ``nominal'' values for the Au+Au 
total cross-section ($\sigma_{AuAu} = 6.85$ b) and BBC-LL1 trigger efficiency 
($\epsilon_{BBC}$ = 92 $\pm$ 3\%)~\cite{ppg014}.

%
\subsection{Trigger and data sample}

The events used in this analysis were collected by a level-1 UltraPeripheral (LL1-UPC)
trigger set up for the first time in Run-4, with the following characteristics:
\begin{enumerate}
\item Veto on coincident signals in both BBC (covering $3.0<|\eta|<3.9$ and full azimuth) 
is imposed in order to reduce peripheral nuclear and beam-gas collisions.
\item A large energy ($E>$ 0.8 GeV) cluster in EMCal is required to select the $e^\pm$ 
from the $\jpsi$ decay ($E\approx 0.5\,m_{\jpsi}\approx$ 1.6 GeV) and from the high-mass
di-electron continuum.
\item At least 30 GeV energy deposited in one or both of the ZDCs is required to 
select Au+Au events with forward neutron emission ($Xn$) from the (single or double) 
Au$^\star$ decay.
\end{enumerate}

\noindent
The trigger used to detect high-energy electrons in the central arm was 
the standard EMCal-RICH-Trigger (ERT) with 2$\times$2 tile threshold at 0.8 GeV. 
A software algorithm performed a crude reconstruction of EMCal clusters by
summing the pedestal-subtracted and gain-calibrated energies in overlapping ``tiles''
of 2$\times$2 calorimeter towers with an estimated (preliminary) efficiency for 
high-energy $e^\pm$ coming from the $\jpsi$ decay of 
$\varepsilon_{\ensuremath{\it trigg}}^{\jpsi} \,=\, 0.9 \pm 0.1$.\\

\noindent
The total number of events collected by the UPC trigger was 8.5 M (i.e. 0.4\% of
the min.bias Au+Au trigger) and the data set comprised $\sim$ 1000 raw-data files 
($\sim$1. TB). Standard QA further reduced this number down to 6.7 M events.
Most of these UPC-triggered events were not, however, signals of interest in this
analysis ({\it coherent} $\jpsi$ and high-$m_{inv}$ $e^+e^-$). Other events 
(with their discriminating characteristics indicated) were likely to fire also the
UPC trigger~\cite{baur98,baur01,bertula05}:
\begin{description}
\item (1) Cosmic rays: no ZDC, no good vertex.
\item (2) Beam-gas collisions: no good vertex, large multiplicity, asymmetric $dN/dy$.
\item (3) Peripheral nuclear collisions: comparatively large particle multiplicities, large pair $p_T$.
\item (4) (Hard) diffractive collisions: Pomeron-Pomeron events with rapidity 
gap(s)~\cite{engel96,roldao00} are usually accompanied by forward proton emission, 
$p_T(\pom\pom) > p_T(\gaga)$, and like-sign pairs.
\item (5) Coherent $\gaga \rightarrow q\bar{q}$: produce mainly hadrons 
(removable by the $e^\pm$ identification cuts).
\item (6) Incoherent (or ``quasi-elastic'') photon-nucleon ($\gamma\,N$) collisions: 
$ p_T(\gamma\,\pom) > p_T(\gaga)$, wider and asymmetric $dN/dy$, $Xn\geq 2n$ 
(recoiling nucleon induces nuclear break-up with larger probability than in fully 
coherent interactions)~\cite{strikman05}.
\end{description}
\noindent
Processes (1) and (2) can be considered ``non-physical'' sources of background
and are rejected by simple global event analysis cuts. ``Physical'' processes (3), (4), (5) 
can be removed by offline cuts more easily than the $\gamma\,N$ contribution (6) 
which has experimental signatures very similar to true $\gA$ reactions
(see~\cite{strikman05} and Section \ref{sec:cross_sec}).

%
\subsection{Data analysis}

Charged particle tracking in PHENIX central arms is based on a combinatorial Hough transform 
in the track bend plane, with the polar angle determined by PC1 and the collision vertex 
along the beam direction\cite{mitchell_nim}. The original $z$-vertex of the track was measured 
with 1-cm resolution using a specific method based on PC hits and EMCal clusters since the 
standard procedure based on BBC and ZDC is not applicable for UPC events which, 
by definition, do not have BBC coincidences and often do not have ZDC coincidences. 
Track momenta are measured with a resolution $\delta p/p \approx$ 0.7\%+1.0\%$p$ 
$\approx$ 1.7 -- 2.7\% for the relevant range ($p\sim$ 1 -- 3 GeV/$c$) in this analysis.\\

\noindent
The following global cuts were applied to enhance the sample of genuine $\gamma$-induced events:
\begin{enumerate}
\item A standard offline vertex cut $|vtx_{z}| <$ 30 cm was required to select collisions 
well centered in the fiducial area of the central detectors and to avoid tracks close 
to the magnet poles.
\item The maximum event multiplicity allowed was 15 tracks to suppress the 
contamination of non-UPC (mainly beam-gas and peripheral nuclear) reactions 
that fired the UPC trigger.
\end{enumerate}

\noindent
At variance with standard $\jpsi\rightarrow e^+e^-$ analyses in nuclear Au+Au
reactions~\cite{phnx_jpsi} which have to deal with large particle multiplicities, 
we did not need to apply very strict PID cuts in order to identify electrons 
in the clean UPC environment: 
\begin{enumerate}
\item RICH multiplicity $n_0\geq$2 selects $e^\pm$ which fire 2 or more tubes 
separated by the nominal ring radius.
\item Good CNT--EMCal matching is required for candidate tracks with an associated
EMCal cluster without dead-warn towers within a 2$\times$2 tile.
\item An EMCal cluster energy cut ($E_1 >$ 1 GeV $||$ $E_2 >$ 1 GeV) is applied to 
select candidate $e^\pm$ in the plateau region above the turn-on curve of the ERT 
trigger (with 0.8 GeV threshold).
\end{enumerate}

\noindent
Beyond those global or single-track cuts, an additional ``coherent'' identification 
cut was applied by selecting only those $e^+e^-$ candidates detected in opposite 
arms (arm$_1 \neq$ arm$_2$). Such a cut enhances the sample of back-to-back 
di-electrons with low pair $p_T$ as expected for $\gaga,\,\gA$ production.
Finally, $\jpsi$ were reconstructed by standard invariant mass analysis of all $e^\pm$
that passed the aforementioned analysis cuts. Any remaining background was removed
from the $m_{inv}$ distribution by directly subtracting the wrong sign pairs 
($e^+e^+$ or $e^-e^-$) from the unlike-sign pairs in a {\it bin-to-bin basis}.
The $\jpsi$ yield is extracted directly by adding the number of counts within
$\pm3\sigma$ (roughly 2.7 -- 3.5 GeV/$c^2$) of a Gaussian fit of the experimental 
mass peak.

%

\subsection{Acceptance and efficiency corrections}

\begin{figure}[htbp]
\begin{center}
\includegraphics[height=4.cm,width=10.cm]{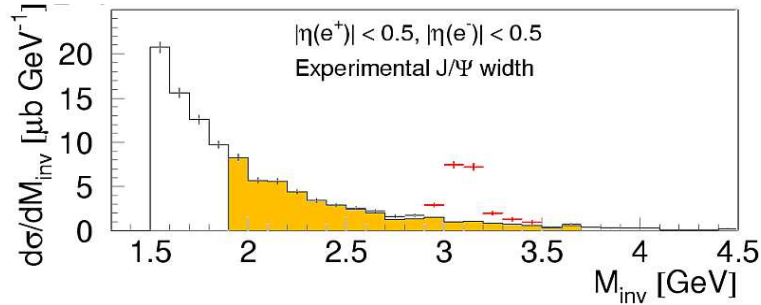}
\end{center}
\vspace{-0.5cm}
\caption{Expected invariant mass distribution of dielectron pairs from 
$\gA\rightarrow \jpsi\rightarrow e^+e^-$ (red points) and 
$\gaga\rightarrow e^+e^-$ (histogram) as given by the {\it Starlight} 
model~\protect\cite{nystrand05} for UPC Au+Au collisions 
at $\sqrtsnn$ = 200 GeV detected in PHENIX. The hatched area indicates the
region accessible with our ERT trigger.}
\label{fig:sim_starlight}
\end{figure}

Physical cross-sections are obtained after correcting the raw number of signal
counts for: (i) the geometrical acceptance of our detector system, and 
(ii) the efficiency losses introduced by the aforementioned analysis cuts.
Acceptance and efficiency corrections were obtained with a 
full Monte Carlo of the experimental apparatus with realistic input distributions
of the physical signals. We generated 10$^5$ events for each one of the two 
coherent process of interest: $\jpsi$ and high-mass $e^+e^-$ pairs production 
in Au+Au collisions accompanied by $Xn$ forward emission, with the 
{\it Starlight} Monte Carlo~\cite{starlight1,starlight2,nystrand05}.
Such a model reproduces well the existing $d^3N/dyd\phi dp_T$ distribution of coherent 
$\rho$ production in UPC Au+Au events measured at RHIC by STAR~\cite{star_rho}.
The coherent events were simulated in the PHENIX detector using GEANT~\cite{geant} 
and passed through the same reconstruction program as the real data.
Figure~\ref{fig:sim_starlight} shows, for illustrative purposes, the 
invariant mass $dN/dm_{ee}$ distributions of the coherent $\jpsi$'s and $e^+e^-$ 
signals given by {\it Starlight} and a {\it fast} Monte Carlo version of our code~\cite{nystrand05}.
A Gaussian $\jpsi$ signal is expected with experimental width of $\sim$100 MeV/$c^2$ 
at $m_{inv}\sim$ 3.1 GeV/$c^2$ on top of a power-law-like $e^+e^-$ continuum.
The $\jpsi$ correction factors obtained from our simulation studies are:
\begin{enumerate}
\item Acceptance ($\jpsi\rightarrow e^+e^-$ decay detected within $|y|<0.5$ and full $\phi$): 
$\ensuremath{{\it Acc}}$ = 5.7\%.
\item Efficiency losses due to all cuts (including yield extraction within $\pm 3\sigma$): 
$\varepsilon_{\ensuremath{{\it reco}}}$ = 56.4\%.
\item Unlike-sign background subtraction results in a negligible efficiency loss
of the $\jpsi$ signal.
\end{enumerate}

\section{EXPERIMENTAL RESULTS}

\begin{figure}[htbp]
\begin{center}
\includegraphics[height=6.8cm,width=6.5cm]{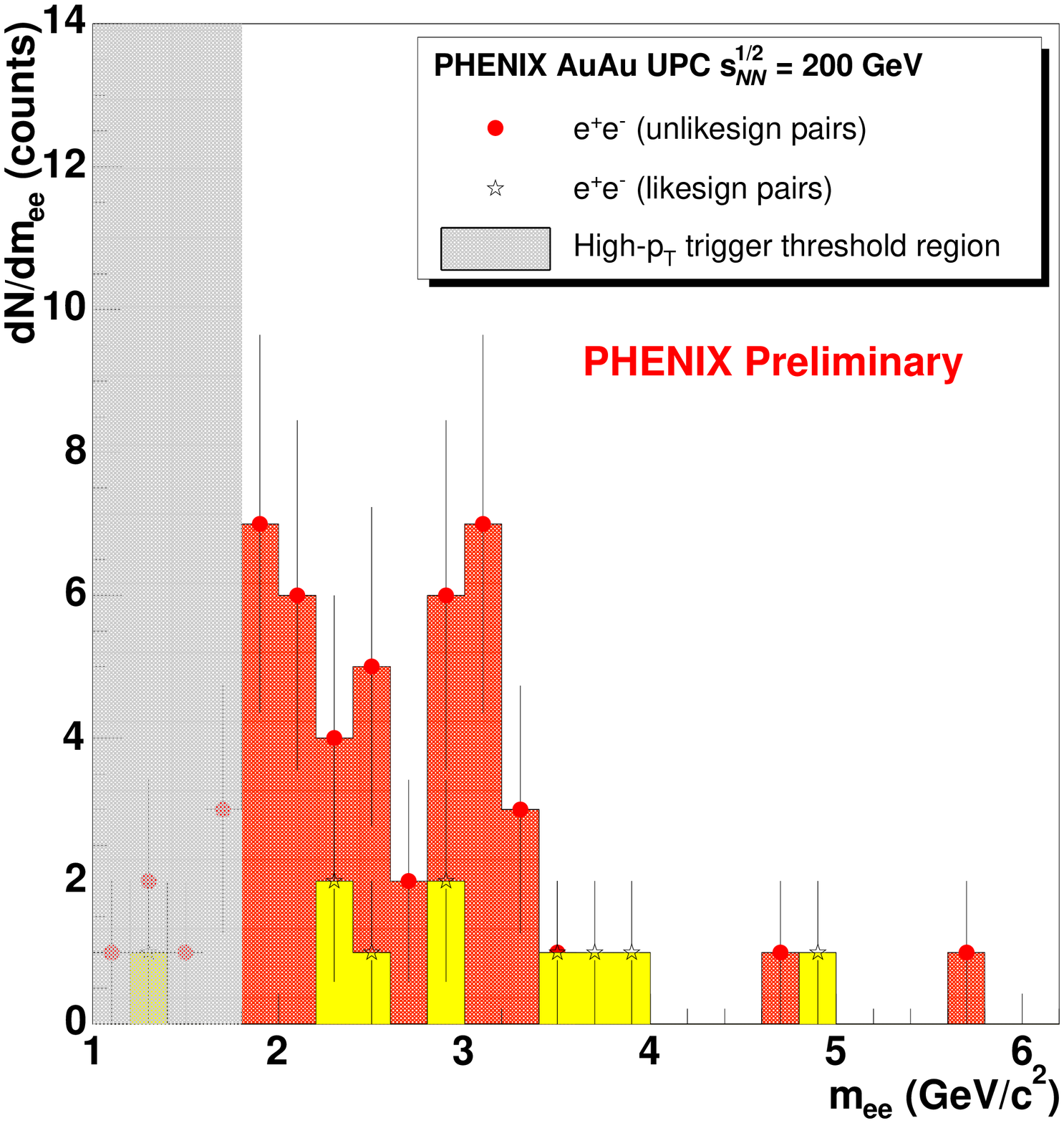}
\includegraphics[height=6.8cm,width=6.5cm]{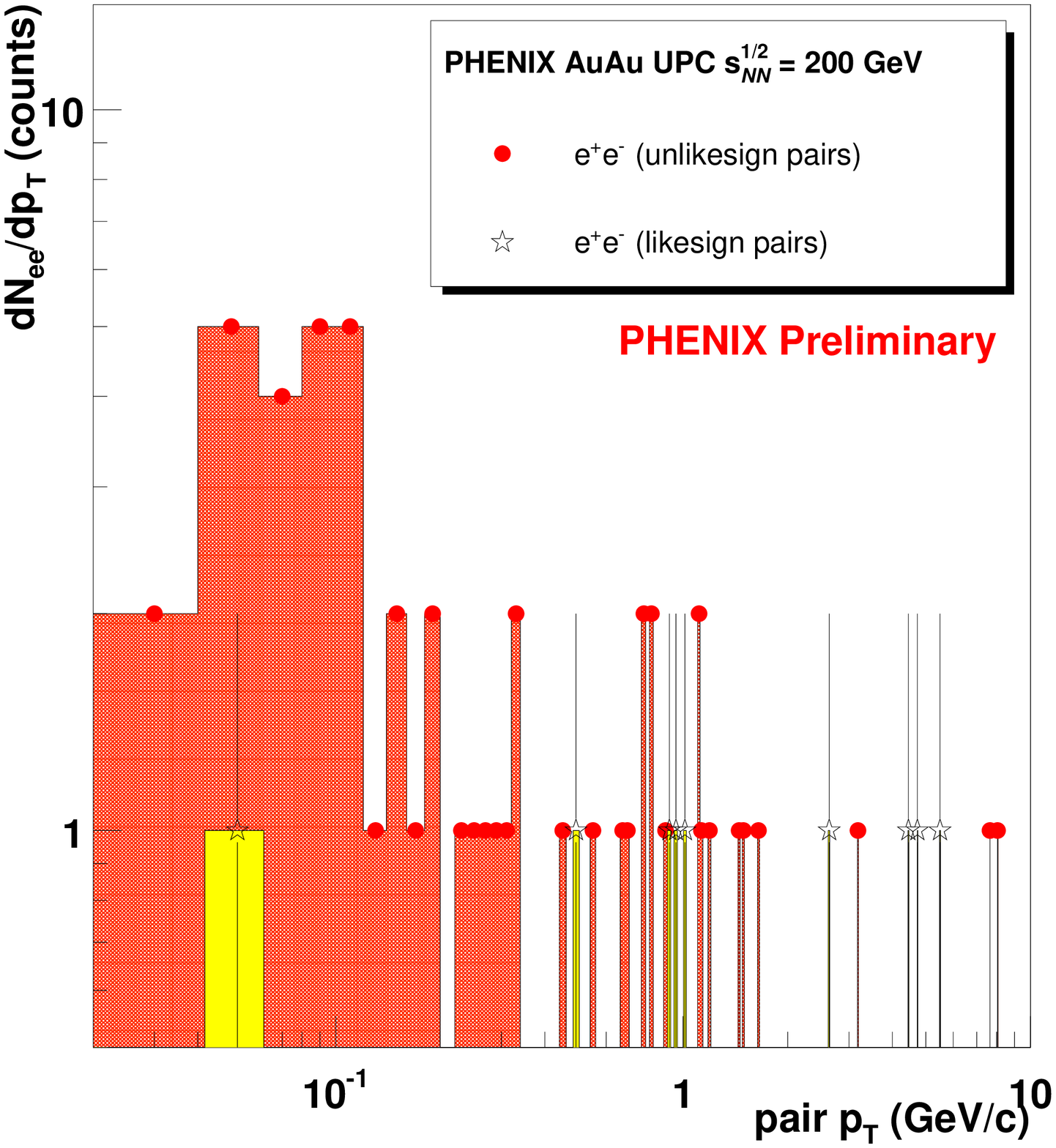}
\end{center}
\vspace{-0.5cm}
\caption{Invariant mass (left) and pair $p_T$ (right) distributions of 
unlike-sign (red dots) and like-sign (open stars) $e^{\pm}e^{\pm}$ pairs identified in 
UPC Au+Au collisions at $\sqrtsnn$ = 200 GeV.}
\label{fig:minv_allpairs}
\end{figure}

Fig.~\ref{fig:minv_allpairs} left, shows the invariant mass of unlike-sign pairs
(red-filled histogram) and same-sign background pairs (yellow-filled 
histo) obtained from our analysis after application of the global, 
single and pair cuts described in the previous section.
The wrong-sign background is very small as expected from the MC results and has 
a flat $p_T$ distribution which extends far from the low-$p_T$ coherent region 
(Fig.~\ref{fig:minv_allpairs} right).
In what follows, all ($m_{inv}$ and $dN/dp_T$) distributions have the like-sign 
background removed. Note that below $m_{inv}\approx$ 1.8 GeV$/c^2$, the applied 
offline cuts above the ERT threshold energy ($E_{1,2}>$ 1 GeV) are responsible 
for the effective lack of counts.

%
\subsection{Invariant mass distribution}

\begin{figure}[htpb]
\begin{center}
\includegraphics[height=7.cm]{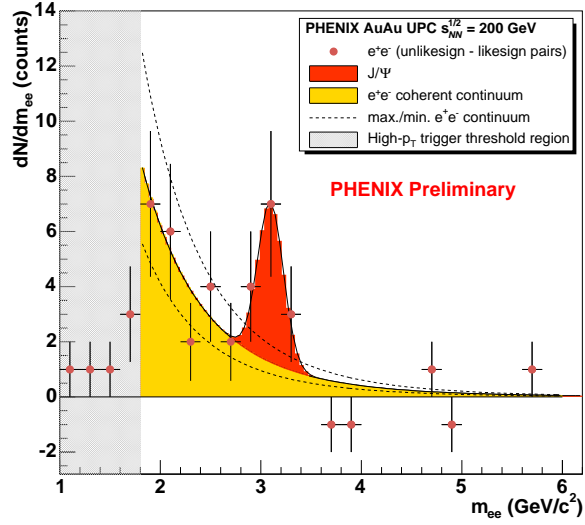}
\end{center}
\vspace{-0.5cm}
\caption{Invariant mass distribution of $e^+e^-$ pairs measured in UPC Au+Au collisions 
at $\sqrtsnn$ = 200 GeV fitted to the combination of a dielectron continuum (power-law 
distribution) and a $\jpsi$ (Gaussian) signal. The two additional dashed curves indicate 
the maximum and minimum continuum contributions considered in this analysis below 
the $\jpsi$ signal region.}
\label{fig:minv_ee_jpsi}
\end{figure}

The final measured background-subtracted $dN/dm_{ee}$ distribution is shown in 
Fig.~\ref{fig:minv_ee_jpsi} including the simulated $e^+e^-$ continuum power-law
curve absolutely normalized at the measured $dN/dm_{ee}$ yield at 
$m_{ee}$ = 2 GeV/$c^2$ combined with a fit to a Gaussian at the $\jpsi$ peak. 
The total number of counts for both physical signals is $N_{e^\pm+\jpsi}\,=\,40 \pm 6$ (stat).
Despite the poor statistics, the shape of the di-electron continuum is consistent 
with the power-law-like distribution expected for the reconstructed shape of the 
$e^+e^-$ MC signal (Fig.~\ref{fig:sim_starlight}). The dotted curves indicate the 
maximum and minimum $e^+e^-$ continuum contributions that we consider in this preliminary 
analysis and have been obtained by absolutely normalizing the power-law curve at the upper 
and lower values given by the $\pm 1\sigma$ yield uncertainty at the 
$m_{ee}$ = 2.0 GeV/$c^2$ bin. From this plot it is obvious that the largest contribution 
to the systematic uncertainty in the extraction of the $\jpsi$ 
signal comes from the subtraction of this physical background.\\

\begin{figure}[htpb]
\begin{center}
\includegraphics[height=7.cm]{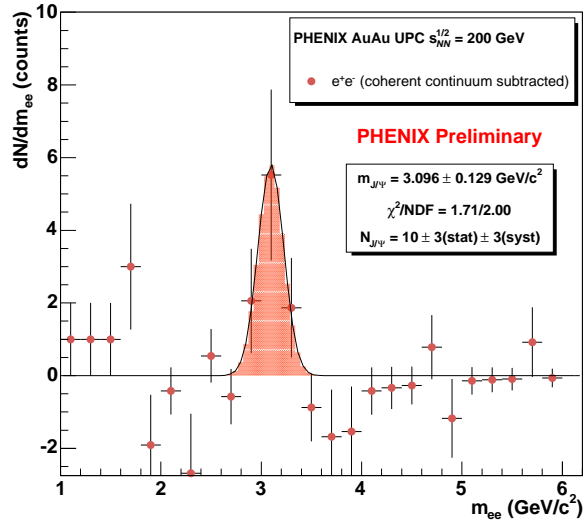}
\end{center}
\vspace{-0.5cm}
\caption{$\jpsi$ invariant mass distribution obtained subtracting from the total $e^+e^-$ pairs signal 
the fitted dielectron continuum shown in Fig.~\protect\ref{fig:minv_ee_jpsi}. The quoted number of
$\jpsi$ is just the number of counts within $m_{ee}$ = 2.7 -- 3.5 GeV/$c^2$ (the error is just
statistical).}
\label{fig:minv_jpsi}
\end{figure}

\noindent
Fig.~\ref{fig:minv_jpsi} shows the resulting invariant mass distribution obtained by 
subtracting the fitted power-law curve of the dielectron continuum from the total 
experimental $e^+e^-$ pairs distribution. There is a clear $\jpsi$ peak whose 
position and width are perfectly consistent with the expected signal from our 
full MC: $m_{\jpsi}$ = 3.096 $\pm$ 130 MeV/$c^2$ (the mass is right at the PDG value, 
though the width seems to be $\sim$30 MeV/$c^2$ larger than the simulated value). 
The total number of $\jpsi$'s within $\pm 3\sigma$ of the peak position 
(2.7 -- 3.5 GeV/$c^2$) is: $N_{\jpsi} \,=\, 10 \pm 3 \mbox{ (stat) } \pm 3 \mbox{ (syst.)}$,
where the systematic uncertainty is completely dominated by the di-electron continuum 
subtraction method. More detailed analyses are being currently carried out to get a 
better handle on the possible $e^\pm$ continuum contamination to the total $\jpsi$ signal.

\subsection{Pair transverse momentum distribution}

\begin{figure}[htpb]
\begin{center}
\includegraphics[height=7.cm]{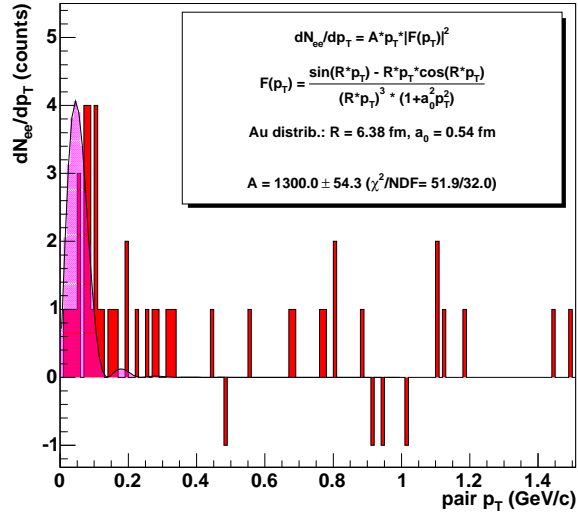}
\end{center}
\vspace{-0.5cm}
\caption{Transverse momentum distribution of reconstructed $e^+e^-$ continuum pairs 
and $\jpsi$ from ultra-peripheral Au+Au collisions compared to the theoretical 
expectations of coherent photoproduction with a realistic nuclear form factor, 
Eq.~(\protect\ref{eq:nucl_form}).}
\label{fig:dNdpT_real_pairs}
\end{figure}

Fig.~\ref{fig:dNdpT_real_pairs} shows the transverse momentum distribution of
all the reconstructed $e^+e^-$ continuum pairs and $\jpsi$ in UPC Au+Au collisions.
Their spectrum is clearly peaked at low $p_T$ as expected for coherent production. 
The {\it shape} of the $p_T$ spectrum itself is well reproduced 
by a nuclear form factor fit~\cite{nystrand01}
\begin{equation}
\frac{dN_{ee}}{dp_{T}} = C\cdot p_{T}\cdot|F(p_{T})|^{2} \;\;\mbox{ with }\;\;
F(p_{T}) = \frac{sin(R\cdot p_{T}) - R\cdot p_{T}\cdot cos(R\cdot p_{T})}{(R\cdot p_{T})^{3} \cdot (1+a_{0}^{2}p_{T}^{2})}
\label{eq:nucl_form}
\end{equation}
with Au nuclear radius and diffusivity fixed to their known values, 
$R$ = 6.38 fm $a_{0}$ = 0.54 fm~\cite{hahn}, and with absolute normalization 
factor $C$ as the only free parameter.

\subsection{Cross-section for coherent UPC $\jpsi$ production at mid-rapidity} 
\label{sec:cross_sec}

\begin{figure}[htbp]
\begin{center}
\includegraphics[height=7.cm]{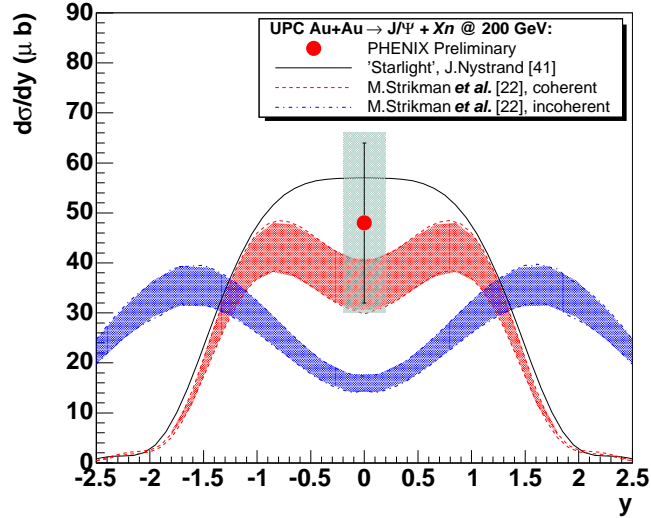}
\end{center}
\vspace{-0.5cm}
\caption{Preliminary cross-section of coherent $\jpsi$ production 
at mid-rapidity in UPC Au+Au collisions at $\sqrtsnn$ = 200 GeV compared to 
the theoretical calculations~\protect\cite{nystrand05,strikman05}. The
error bar (box) shows the statistical (systematical) uncertainties of 
the measurement.}
\label{fig:dNdy_vs_model}
\end{figure}

The final cross-section for coherent $\jpsi$ photoproduction at midrapidity 
in UPC Au+Au collisions at $\sqrtsnn$ = 200 GeV accompanied by Au breakup is:
\begin{eqnarray}
\left .\frac{d\sigma_{\ensuremath{\it UPC}\;\jpsi}}{dy}\right|_{|y|<0.5} =  
\frac{1}{\ensuremath{{\it BR}}}\cdot\frac{N_{\jpsi}}{\ensuremath{{\it Acc}}\cdot\varepsilon_{reco}\cdot
\varepsilon_{\ensuremath{\it trigg}}\cdot {\cal L}_{int}}\cdot\frac{1}{\Delta y}
= 48. \pm 14. \mbox{ (stat)} \pm 16. \mbox{ (syst) } \mu b.
\nonumber
\end{eqnarray}
\noindent
where all correction factors (and corresponding uncertainties) have been obtained
as described in previous sections, and BR = 5.93\% is the known $\jpsi$ 
di-electron branching ratio. The final $\jpsi$ cross-section 
is in very good agreement, within the (still large) experimental errors,
with the theoretical values computed in~\cite{nystrand05,strikman05} as shown in 
Fig.~\ref{fig:dNdy_vs_model} (the predictions of~\cite{strikman05} have been 
scaled down by the nuclear breakup probability $P_{Xn}\sim$ 0.64). 
The current uncertainties preclude yet any detailed
conclusion regarding the two crucial ingredients of the models (nuclear gluon 
distribution and $\jpsi$ nuclear absorption cross-section). Whereas the systematical
uncertainty of the measurement is linked to the (yet) imprecise contribution
from coherent $\gaga\rightarrow e^+e^-$ pairs below the $\jpsi$ peak, the statistical 
error can only be improved with a (much) higher luminosity Au+Au run. The contribution 
of the incoherent $\gamma\,N$ production to the total 
$\left .d\sigma_{\ensuremath{\it UPC}\;\jpsi}/dy\right|_{|y|<0.5}$ (included in
the predictions of Strikman {\it et al.}~\cite{strikman05} but absent from
{\it Starlight}~\cite{nystrand05}) needs to be elucidated too. 

%
\section{CONCLUSION}
\label{sec:}

We have presented preliminary PHENIX results from an analysis of 
Au+Au ($\sqrtsnn$ = 200 GeV) UPC triggered events aiming at the measurement 
of coherent photoproduction of $\jpsi$ and high-mass $e^+e^-$ pairs in 
$\gA$ and $\gaga$ processes respectively. Ultraperipheral Au+Au collisions are
tagged by forward (ZDC) neutron detection from the (single or double) Au$^\star$ 
dissociation. Clear indications of $\jpsi$ and high mass dielectron continuum 
have been found in the data. The total number of ``physics'' $e^+e^-$ pairs is: 
$dN_{e^\pm+\jpsi}/dy$ = 40 $\pm$ 6 (stat). Their $p_T$ spectrum is peaked at 
low $p_T\approx$ 90 MeV/$c$ as expected for coherent photoproduction
with a realistic nuclear form factor.
After subtraction of the physical $e^+e^-$ signal, the measured invariant mass 
distribution has a clear peak at the $\jpsi$ mass with experimental width in 
good agreement with a full GEANT-based simulation for UPC production and 
reconstruction in PHENIX. The measured number of $\jpsi$'s in PHENIX acceptance is: 
$dN/dy$ = 10 $\pm$ 3 (stat) $\pm$ 3 (syst). After correcting for acceptance and 
efficiency losses and normalizing by the measured luminosity, we obtain a 
preliminary cross-section of $d\sigma/dy|_{y=0}$ = 48 $\pm$ 16 (stat) $\pm$ 18 (syst) $\mu b$ 
which is consistent, within errors, with theoretical expectations. 
Foreseen improvements in our analysis will likely reduce the experimental uncertainties 
of the measured cross-section and provide more quantitative information on the nuclear 
gluon distribution and $\jpsi$ absorption in cold nuclear matter at RHIC 
energies.

\begin{ack}
\noindent
This work is the result of a collaborative effort of the UPC PHENIX crew
(M.~Chiu, D.d'E., J.~Nystrand, D.~Silvermyr and S.~White). Useful discussions 
with L.~Frankfurt, S.~Klein, M.~Strikman, and M.~Zhalov are acknowledged.

\end{ack}

\clearpage


\def\IJMPA{{Int. J. Mod. Phys.}~{\bf A}}
\def\EPJ{{Eur. Phys. J.}~{\bf C}}
\def\JPG{{J. Phys}~{\bf G}}
\def\JHEP{{J. High Energy Phys.}~}
\def\NCA{Nuovo Cimento~}
\def\NIM{Nucl. Instrum. Methods~}
\def\NIMA{{Nucl. Instrum. Methods}~{\bf A}}
\def\NPA{{Nucl. Phys.}~{\bf A}}
\def\NPB{{Nucl. Phys.}~{\bf B}}
\def\PLB{{Phys. Lett.}~{\bf B}}
\def\PLC{Phys. Rept.\ }
\def\PR{Phys. Rev.\ }
\def\PRL{Phys. Rev. Lett.\ }
\def\PRD{{Phys. Rev.}~{\bf D}}
\def\PRC{{Phys. Rev.}~{\bf C}}
\def\ZPC{{Z. Phys.}~{\bf C}}


\end{document}